\documentclass[aip,pop,reprint]{revtex4-1}

\usepackage{graphicx}                       
\usepackage[outdir=./]{epstopdf}
\usepackage{amsmath}
\usepackage{amssymb}
\usepackage{amsthm}
\usepackage{dsfont}       
\usepackage[usenames,dvipsnames,table]{xcolor}
\usepackage[format=plain, font=small, labelfont=bf, justification = raggedright]{caption}
\usepackage{array}  
\usepackage{overpic}
\usepackage{placeins} 
\usepackage{fancyhdr}
\usepackage{colortbl}
\usepackage{multirow} 
\usepackage{hyperref} 
\hypersetup{colorlinks = true,linkcolor = blue, breaklinks = true}

\newcommand{\eq}[1]{(\ref{#1})}
\newcommand{\Eq}[1]{Eq.~\eq{#1}}
\newcommand{\Eqs}[1]{Eqs.~\eq{#1}}
\newcommand{\Fig}[1]{Fig.~\ref{#1}}
\newcommand{\Sec}[1]{Sec.~\ref{#1}}

\newcommand{\Refs}[1]{Refs.~\onlinecite{#1}}
\newcommand{\App}[1]{Appendix~\ref{#1}}

\newcommand{\eg}{{e.g., }}
\newcommand{\ie}{{i.e., }}

\newcommand{\mc}[1]{\mathcal{#1}}

\newcommand{\pd}[1]{\partial_{#1}}

\newcommand{\dd}{\mathrm{d}}

\DeclareMathOperator{\airyA}{Ai}

\newcommand{\Vect}[1]{{\boldsymbol{\rm #1}}}

\newcommand{\Symb}[1]{\mc{#1}}

\renewcommand{\Im}{\textrm{Im}}

\newcommand{\nullFrac}{\vphantom{\frac{}{}}}

\newcommand{\Stroke}[1]{\text{\ooalign{ $#1$\cr \hidewidth\raise.225ex \hbox{$-\mkern.5mu$}\cr}}}

\begin{document}
\setlength{\parskip}{0pt}
\setlength{\belowcaptionskip}{0pt}


\title{Regarding the extension of metaplectic geometrical optics to modelling evanescent waves in ray-tracing codes}
\author{N. A. Lopez}
\email[corresponding author: ]{nicolas.lopez@physics.ox.ac.uk}
\affiliation{Rudolf Peierls Centre for Theoretical Physics, University of Oxford,
Oxford OX1 3PU, UK}
\author{R. H\o jlund}
\affiliation{Section for Plasma Physics and Fusion Energy, Department of Physics,
Technical University of Denmark, DK-2800 Kgs. Lyngby, Denmark}
\author{M. G. Senstius}
\affiliation{Rudolf Peierls Centre for Theoretical Physics, University of Oxford,
Oxford OX1 3PU, UK}

\begin{abstract}

Metaplectic geometrical optics (MGO) is a recently developed ray-tracing framework to accurately compute the wavefield behavior near a caustic (turning point or focal point), where traditional ray-tracing breaks down. However, MGO has thus far been restricted to having real-valued wavevectors. This is disadvantageous because often upon crossing a caustic from the `illuminated' region to the `shadow' region, two real-valued rays coalesce into one complex-valued ray corresponding to the transition from propagating to evanescent behavior. One can distinguish caustics as having either `illuminated shadows' or `proper shadows' -- the former corresponds to when the shadow still contains real-valued rays (albeit in a fewer quantity than in the illuminated region), while the latter corresponds to when the shadow contains no real-valued rays. Here, by means of examples, we show how MGO can be used to model both types of shadows. First, for illuminated shadows we show that MGO can actually be used `as is', provided a corrected integration scheme is used compared to that proposed in the original references. This is then implemented and demonstrated in a recently developed MGO ray-tracing code. Second, we show that for proper shadows, the MGO formalism can still be used if the symplectic rotation matrix that removes caustics along rays is allowed to be complex-valued. In both cases, strong agreement is seen between the MGO and the exact solution, demonstrating the potential of MGO for improving the predictive capability of ray-tracing codes and laying the foundations for modeling more complicated evanescent phenomena such as tunnelling with MGO.

\end{abstract}

\maketitle

\pagestyle{fancy}
\lhead{Lopez, H\o jlund, \& Senstius}
\rhead{MGO for evanescent waves}
\thispagestyle{empty}

\section{Introduction}

Next-generation fusion devices such as STEP~\cite{Freethy23,Tholerus24} and ITER~\cite{Henderson15} rely heavily on externally launched electromagnetic waves to provide heating and current drive. Modeling the performance of these waves is often done using ray-tracing codes, \eg GENRAY~\cite{Smirnov03}. However, the ray-tracing formalism breaks down at turning points and focal points (generally known as caustics), which are commonly encountered as a wave propagates in inhomogeneous plasma. Developing the means of modeling caustics within ray-tracing codes, especially in an emergent manner when the caustic details (\ie type, location) are not known a priori, would greatly improve the predictive capability of such codes.

To address this issue, a new ray-tracing formalism called metaplectic geometrical optics (MGO) was developed~\cite{Lopez19,Lopez20,Lopez21a,Donnelly21,Lopez22} to reinstate the ray approximation at caustics, which has also been implemented in a proof-of-principle one-dimensional (1D) code~\cite{Hojlund24}. However, the MGO formalism is based on real-valued ray-tracing in which the local wavevector of the wavefield is always real. This means that MGO is currently unable to model evanescent decaying fields whose local wavevector is imaginary. Consider an X-polarized electromagnetic wave launched into a high-density magnetized plasma. Such a wave will encounter the right-hand cutoff, transform into an evanescent wave, then tunnel to the upper-hybrid resonance. In steady-state, a standing-wave pattern with complete transmission of the incident X-mode wave across the evanescent region can be obtained if the distance between the cutoff and resonance is not too large~\cite{Ram96,Lopez18b}. Such a scenario might be useful for heating and driving current in spherical tokamaks~\cite{Ram00,Ram05}, but is unable to be modeled via MGO due to the lack of complex rays.

Here we undergo two case studies to develop understanding for how MGO can be used to model evanescent waves. First, we study a one-dimensional analogue of a cusp caustic. (The cusp caustic commonly occurs near focal points.) This example is chosen because the exact solution contains the superposition of one propagating wave (corresponding to a real-valued ray) and one evanescent wave (corresponding to a complex-valued ray). We show that, although the complex ray is not formally included within MGO, in practice it is possible to obtain its contribution in a numerical implementation because once the complex ray becomes close enough to the real line (meaning the magnitude of the evanescent wave becomes comparable to the propagating one), it becomes numerically degenerate with the real-valued ray. This principle is demonstrated by showing that a recently developed MGO code~\cite{Hojlund24} can be brought into strong agreement with the exact solution, even though no complex rays are formally included. As a prerequisite result to this, we also correct an error in the quadrature rule proposed originally in Ref.~\onlinecite{Donnelly21}, and show that the corrected method enables less-supervised simulations.

Second, we study the prototypical fold caustic described by Airy's equation (physically modeling reflections), with the specific aim of extending the MGO solution beyond the turning point into the evanescent region. Since there are no real-valued rays there, the standard MGO formalism does not work. However, we show that by analytically continuing the MGO formalism via upgrading the symplectic rotation matrices to be complex-symplectic instead, the obtained solution is nearly identical with the exact result. Thus, in both cases, the inclusion of complex rays within MGO is shown to be simpler than anticipated. This continues to demonstrate the potential utility of using MGO to develop improved ray-tracing codes.

This paper is organized as follows: in \Sec{sec:review} the basic equations of MGO are reviewed; in \Sec{sec:cusp} it is demonstrated how MGO can be used to model the `illuminated shadow' of a cusp caustic to reasonable accuracy even though no complex rays are formally included; in \Sec{sec:airy} a new MGO formalism is proposed to model evanescent waves and is validated by reproducing the exponential decay of the Airy function to remarkably high accuracy; finally, in \Sec{sec:concl} the main results are summarized. Appendices present auxiliary results, namely, a simplified MGO formalism (\App{app:simpleMGO}) and the corrected quadrature rule (\App{app:correctANGLE}).

\section{Review of MGO}
\label{sec:review}

Here we briefly summarize the MGO formalism in 1D. For detailed derivations of these results, the reader is invited to consult Ref.~\onlinecite{Lopez22t}. Also, a simplified version of MGO for semi-analytical computations is presented in \App{app:simpleMGO}.

Fundamentally, MGO is a ray-based theory, so as with traditional ray-tracing methods~\cite{Tracy14}, one begins by tracing rays via Hamilton's equations of motion
\begin{equation}
	\dot{x}(\tau) = \pd{k} \Symb{D}
	, \quad
	\dot{k} = - \pd{x} \Symb{D}
	,
	\label{eq:rays}
\end{equation}

\noindent where $\Symb{D}$ is the appropriate dispersion relation for the wave of interest, and $x$ and $k$ are the position and local wavevector of the wavefield at a given time $\tau$. In doing so, one obtains the solution set $(x(\tau), k(\tau))$. Although the solution set is single-valued (non-intersecting) when viewed in the $(x,k)$ phase-space, the individual projections $x(\tau)$ or $k(\tau)$ are not. This means that at a given spatial position $x$ there can be multiple rays that contribute to the solution. Hence, the solution for a wavefield $E(x)$ from any ray-based method will be given generically by a sum over individual ray contributions as
\begin{equation}
	E(x) = \sum_{\text{rays } j} \psi_j(x)
	,
    \label{eq:sum}
\end{equation}

\noindent where $\psi_j(x)$ is the contribution from the $j$-th ray at location $x$.

In MGO, the $j$-th ray contribution takes the following form~\cite{Lopez22}: for $\dot{k}(\tau_j) = 0$ one has
\begin{subequations}
	\begin{equation}
		\psi_j(x)
		=
		c
		\sqrt{ 
			\frac{
				\dot{x}(0) 
			}{
				\dot{x}(\tau_j)
			} 
		}
		\exp\left[
			i \int_0^{\tau_j}
			\dd t \,
			k(t) \dot{x}(t)
		\right], 
		\label{eq:MGOk0}
	\end{equation}

	\noindent which is equivalent to the standard ray-tracing expression, or when $\dot{k}(\tau_j) \neq 0$ one has%
	\footnote{Note that this step tacitly assumes $x$ and $k$ have been appropriately normalized and are therefore dimensionless.}%
	\begin{align}
		\psi_j(x)
		&=
		c
		\sqrt{ 
			\frac{
				i \, \dot{x}(0) 
			}{
				2\pi \dot{k}(\tau_j)
			} 
		}
		\exp\left[
			i \int_0^{\tau_j}
			\dd t \,
			k(t) \dot{x}(t)
		\right]
		\nonumber\\
		&\times
		\int_{\mc{C}_0} \dd \epsilon \,
		\Psi_j(\epsilon)
		\exp\left[
			- \frac{
				i \dot{x}(\tau_j)
			}{
				2 \dot{k}(\tau_j)
			} \epsilon^2
			- i K(\tau_j) \epsilon
		\right]
        .
		\label{eq:MGOk!0}
	\end{align}
	\label{eq:MGO}
\end{subequations}

\noindent In either case, $c$ is a constant fixed by boundary conditions and $\tau_j$ is the time duration needed for the $j$-th ray to reach the observation point $x$ from the initial plane encoded by the ray location $x(0)$. Furthermore, in the second expression, $\Psi_j$ is the standard ray-tracing contribution \eq{eq:MGOk0} of the $j$-th ray, but obtained in a symplectically transformed reference frame~\cite{Lopez20}. For brevity, we shall not list the explicit form of $\Psi_j$, and instead refer the reader to Refs.~\onlinecite{Lopez22, Lopez22t, Hojlund24} for details. Also, $K(\tau_j)$ corresponds to the symplectically transformed wavevector, obtained from the original rays as
\begin{equation}
	K(\tau_j) =  
	\frac{
		\dot{x}(\tau_j) k(\tau_j) 
		- \dot{k}(\tau_j) x(\tau_j)
	}{
		\sqrt{ [\dot{x}(\tau_j)]^2 + [\dot{k}(\tau_j)]^2 }
	} 
	.
    \label{eq:rotK}
\end{equation}


\noindent Lastly, $\mc{C}_0$ denotes the steepest-descent contour that passes through the saddlepoint at $\epsilon = 0$. More is said about $\mc{C}_0$ in \App{app:correctANGLE}, which also describes the corrected quadrature rule for numerical implementations of MGO. In \Eqs{eq:MGO} and \eq{eq:rotK}, all of $x$, $k$, and $\tau$ are real-valued, but this constraint will be re-evaluated in the following sections.

On this note, it is worth remarking that the new quadrature rule described in \App{app:correctANGLE} allows complex saddlepoints (\ie complex rays) to be included in the MGO solution \eq{eq:MGOk!0} when they become most important, even though they are formally absent. Complex saddlepoints become most important when they begin to coalesce with real saddlepoints, otherwise they are exponentially damped~\cite{Wright80}. Although formally this coalescence only occurs exactly at a caustic, numerically it will occur in a small neighborhood of the caustic. When that happens, the new method is able to include this additional contribution by effectively integrating along the union of the two individual steepest-descent contours. This is seen in \Fig{fig:contEVO} for the cusp caustic: the steepest-descent valleys tracked by the new method actually correspond to the valleys of the nearby complex saddlepoint, not the real saddlepoint at $\epsilon = 0$. When the two saddlepoints become sufficiently close [determined by details of the root-finding procedure \eq{eq:PHIfit}], the new algorithm treats them as effectively coalesced. The result is a smoother field than one would obtain if the complex saddlepoints were omitted entirely. 

This implies that, at least in the near term, upgrading the MGO formalism to include complex rays is not as necessary as hypothesized~\cite{Lopez21a} for modeling `illuminated shadows' that contain both real and complex rays. This point shall now be demonstrated in the following section.

\section{Modeling `illuminated shadows' with standard MGO: Cusp caustic case study}
\label{sec:cusp}

In Ref.~\onlinecite{Lopez21a}, the paraxial cusp caustic was analyzed using MGO in two spatial dimensions. Being in 2-D, however, led to complicated expressions and prevented a clear understanding of how accurate MGO actually is for the cusp caustic, particularly regarding how the lack of complex rays might limit the accuracy of MGO. Here we analyze a 1-D cusp caustic instead that turns out to be exactly solvable within the MGO formalism. This ultimately enables us to draw concrete conclusions regarding the utility of MGO for modeling `illuminated shadows', that is, the field pattern near a caustic that results from interference between both real and complex rays.

\subsection{Analytical expressions}

Let us consider a wave that propagates according to the equation%
\footnote{Equation \eq{eq:cuspEQ} can be modified by adding a term $i a \pd{x} E$ to observe the unfolding of the cusp caustic, but we shall not consider this.}%
\begin{equation}
	i \pd{x}^3 E(x) + x E(x) = 0
	.
	\label{eq:cuspEQ}
\end{equation}

\noindent Although we introduce this equation simply as an academic exercise to study the cusp caustic in its simplest formulation, physically \Eq{eq:cuspEQ} also describes the evolution of the spectral content of a wavefield propagating paraxially in a cubic potential. 

\begin{figure}[t!]
	\includegraphics[width=0.8\linewidth,trim = {3mm 3mm 3mm 3mm}, clip]{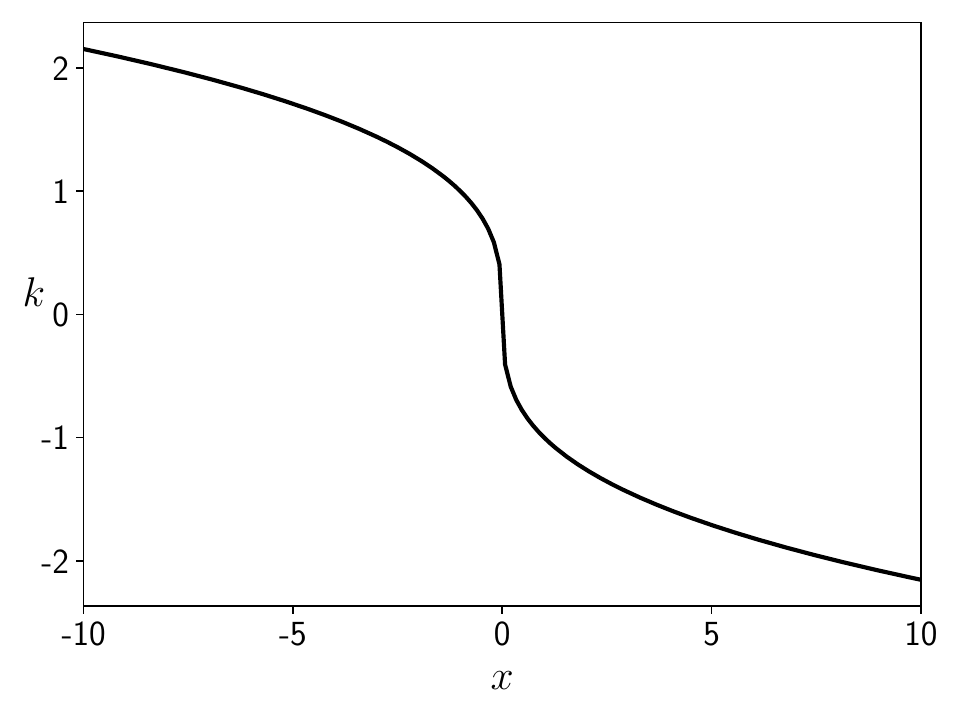}
	\caption{Ray trajectory given by \Eq{eq:cuspRAY} corresponding to the wave equation provided by \Eq{eq:cuspEQ}. A cusp caustic occurs at $x = 0$ where the trajectory has an inflection point.}
	\label{fig:cuspRAY}
\end{figure}

Through use of the Fourier transform, it can be shown that the exact (bounded) solution
\footnote{Fourier methods naturally select the only bounded solution to \Eq{eq:cuspEQ} since it is the only one that possesses a well-defined Fourier Transform. The same is true for \Eq{eq:AiryEQ}.}
to \Eq{eq:cuspEQ} is
\begin{equation}
	E_\text{ex}(x)
	= \textrm{Pe}(\sqrt{2} \, x, 0)
	,
	\label{eq:EXcusp}
\end{equation}

\noindent where $\textrm{Pe}$ is the Pearcey function~\cite{Olver10a}, which has the following equivalent integral representations~\cite{Paris91}:
\begin{align}
	&\textrm{Pe}(x, y)
	= \int_{-\infty}^\infty \dd \kappa  \exp\left( i \kappa^4 + i y \kappa^2 + i \kappa x \right)
	\nonumber\\
	&=
	2 e^{i \pi/8}
	\int_0^\infty \dd t \,
	\exp\left( - t^4 - y t^2 e^{-i \pi/4} \right) \cos\left( x t e^{i\pi/8} \right)
	.
    \label{eq:pearceyINT}
\end{align}

\noindent As is well-known~\cite{Wright80,Paris91}, the evaluation of \Eq{eq:pearceyINT} when $y = 0$ requires the computation of two saddlepoint contributions within the complex $\kappa$ plane, one real-valued and one complex-valued. These two saddlepoints coalesce when $x = 0$ to form a degenerate saddle at $\kappa = 0$. Physically, this means that the field given by \Eq{eq:EXcusp} results from the interference between one real-valued ray (corresponding to a propagating wave) and one complex-valued ray (corresponding to an evanescent decaying wave). It is for this feature that \Eq{eq:cuspEQ} was chosen as a case study for modeling `illuminated shadows' with MGO.

The dispersion relation corresponding to \Eq{eq:cuspEQ} is
\begin{equation}
	\Symb{D} = k^3 + x = 0
	.
\end{equation}

\noindent Correspondingly, the ray trajectory that solves \Eq{eq:rays} is given as
\begin{equation}
	x(\tau) = \tau^3
	, \quad
	k(\tau) = - \tau
	.
	\label{eq:cuspRAY}
\end{equation}

\noindent The ray trajectory \eq{eq:cuspRAY} is plotted in \Fig{fig:cuspRAY}. Importantly, note that for every value of $x$ there is only one ray; hence there is no need to sum over rays in \Eq{eq:sum}.

Since one has $\dot{x} = 3 \tau^2$, the standard GO solution \eq{eq:MGOk0} is calculated to be
\begin{equation}
	E_\text{GO}(x)
	= 
	\sqrt{\pi}
	\frac{
		\exp\left( - i \frac{3}{4} |x|^{4/3} + i \frac{\pi}{4}\right)
	}{
		\sqrt{3 |x|^{2/3}} 
	}
	,
	\label{eq:GOcusp}
\end{equation}

\noindent where we have set $c$ in accordance with the known result~\cite{Olver10a}, and have inverted \Eq{eq:cuspRAY} to obtain $\tau_j = x^{1/3}$.

To compute the MGO solution via \Eq{eq:MGOk!0}, we first note that $\dot{k} = -1$. Then, using the formalism outlined in Refs~\onlinecite{Lopez22} one can compute $\Psi_j(\epsilon) \equiv \Psi(\epsilon; \tau_j)$ as
\begin{widetext}
\begin{align}
    \Psi(\epsilon; \tau_j)
    =  \frac{
        \vartheta(\tau_j)
        \exp\left\{
            i \Theta[ \vartheta(\tau_j) \epsilon; \tau_j] 
            - i \Theta(0; \tau_j)
            - i \frac{2 t^3}{\vartheta(t)} \epsilon
            - i \frac{3}{2} t^2 \epsilon^2
            \nullFrac
        \right\}
    }{
        \sqrt{ 
            9 \tau_j^2 \left\{
                T[ \vartheta(\tau_j) \epsilon; \tau_j \nullFrac]
            \right\}^2 
            + 1 
        }
	}
    ,
\end{align}

\noindent where $\vartheta(t) = \sqrt{ 1 + 9 t^4}$, the phase function is given by
\begin{align}
    \Theta(\varsigma; t)
    &=
    \frac{\varsigma^2}{6 t^2}
	+ \frac{\varsigma}{3 t}
    +
    3 \frac{
        4
        + (27 t^6 + 9 t^2 + 9 t\varsigma)^2
    }{648 t^4} \,
    \pd{\varsigma} T( \varsigma; t )
    - \frac{9 t^5 + 3t + 3\varsigma}{8 t^2} \, T( \varsigma; t )
    ,
\end{align}

\noindent and we have introduced the auxiliary function (which one might recognize as the rotated ray map)
\begin{align}
    T(\varsigma; t)
    =
	\frac{
		\left[
			3 t^5 + t 
            + \varsigma
			+ \sqrt{
				\frac{4}{81 t^2}
				+ (3 t^5 + t + \varsigma)^2
			}
		\right]^{1/3}
	}{
		(6 t^2)^{1/3}
	}
    +\frac{
		\left[
			3 t^5 + t 
            + \varsigma
			- \sqrt{
				\frac{4}{81 t^2}
				+ (3 t^5 + t + \varsigma)^2
			}
		\right]^{1/3}
	}{
		(6 t^2)^{1/3}
	}
 .
\end{align}

\end{widetext}

The MGO solution \eq{eq:MGOk!0} is therefore given as
\begin{align}
    &E_\text{MGO}(x)
    = 
	\frac{
		\exp\left( - i \frac{3}{4} |x|^{4/3}\right)
	}{
		\sqrt{2} 
	}
 \nonumber\\
 &\times
    \left.
    \int_{\mc{C}_0} \dd \epsilon \,
    \Psi(\epsilon; \tau_j)
    \exp\left[
        i \frac{2 \tau_j^3}{\vartheta(\tau_j)} \epsilon
        + i \frac{3}{2} \tau_j^2 \epsilon^2
        \nullFrac
    \right]
    \right|_{\tau_j = x^{1/3}}
    .
    \label{eq:MGOcusp}
\end{align}

\noindent Note that we have used the same value for the overall constant in MGO as in standard GO, as must be the case since the two methods are asymptotically equivalent far from a caustic~\cite{Lopez22}. Also note that when evaluated at the caustic $x = 0$, one has $E_\text{GO}(0) \to \infty$ while
\begin{align}
	E_\text{MGO}(0)
	&=
	\frac{1}{\sqrt{2}}
	\int_{\mc{C}_0}
	\dd \epsilon \,
	\exp\left(
		\frac{
			i
		}{
			4
		} \epsilon^4
	\right)
	\nonumber\\
	&=
	\int_{-\infty}^\infty
	\dd \kappa \,
	\exp\left(
		i \kappa^4
	\right)
	\nonumber\\
	&=
	\textrm{Pe}(0, 0)
	\equiv E_\text{ex}(0)
	,
\end{align}

\noindent where we have deformed the steepest-descent contour back along the real line, since there are no additional saddlepoints or poles in the integrand. In other words, the MGO solution gets the exact answer at the caustic%
\footnote{This is actually expected because at the caustic, the tangent plane is parallel to the $k$-axis. The metaplectic transform is then simply a Fourier transform, same as the exact result.}.%

\subsection{Numerical results}

Now let us evaluate the MGO solution \eq{eq:MGOcusp} numerically using a tenth-order Gauss--Freud quadrature rule (modified according to \App{app:correctANGLE}). The computation is performed from $x = -10$ to $x = 10$ and crosses over the caustic at $x = 0$. No attempt is made to prescribe the caustic location at $x = 0$ or impose that the solution be symmetric about $x = 0$.

\begin{figure}
    \centering
    \includegraphics[width=\linewidth,trim={3mm 3mm 3mm 3mm},clip]{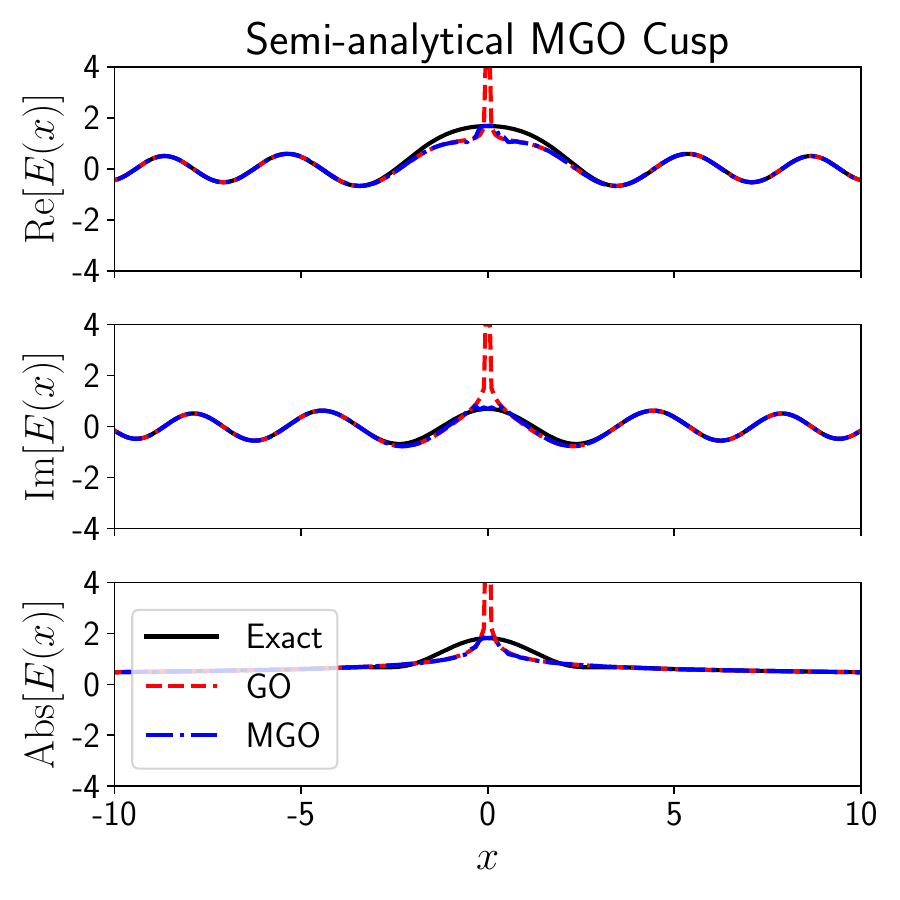}
    \caption{Comparison of the MGO solution for the 1D cusp caustic given by \Eq{eq:MGOcusp} with the standard GO solution given by \Eq{eq:GOcusp} and the exact solution given by \Eq{eq:EXcusp}.}
    \label{fig:cusp_new}
\end{figure}

The results of this semi-analytical computation are shown in \Fig{fig:cusp_new}. The solution remains finite at the caustic as desired, and remains well-behaved as the simulation progresses beyond the caustic. Note also that the exact parity of the solution is preserved, because the exact expressions that enter into \Eq{eq:MGOcusp} also obey this parity.

\begin{figure}
    \centering
    \includegraphics[width=\linewidth,trim={3mm 3mm 3mm 3mm},clip]{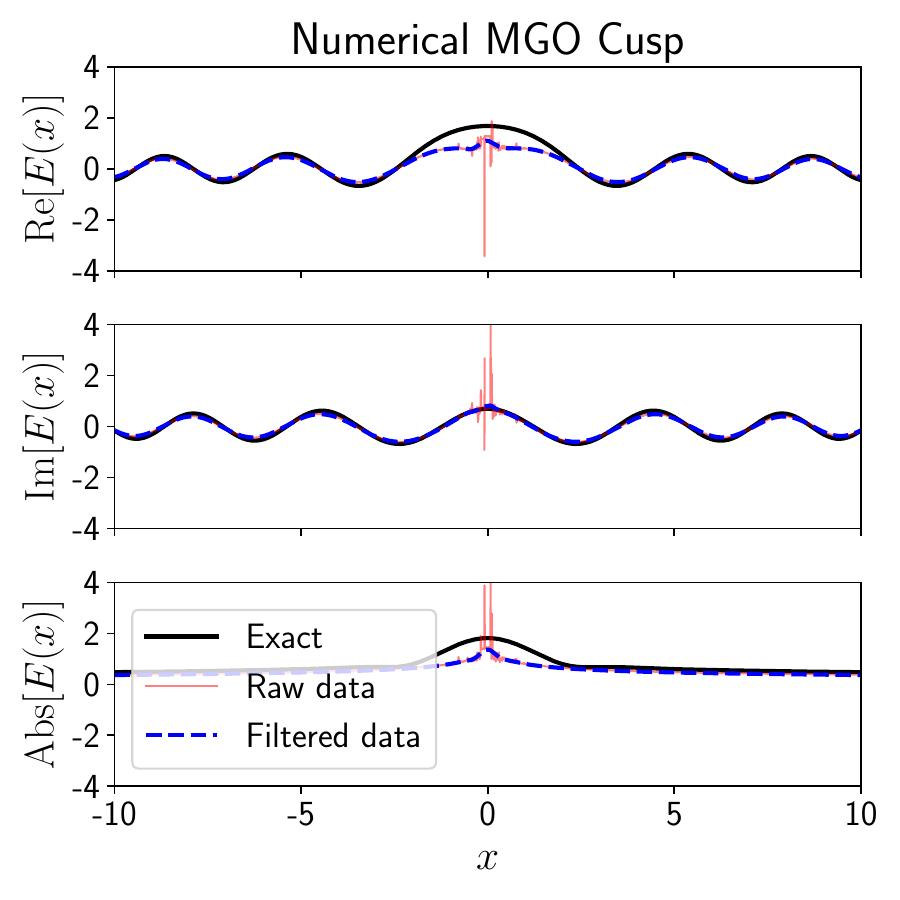}
    \caption{Comparison of the exact solution for the 1D cusp caustic given by \Eq{eq:EXcusp} with the solution provided by the MGO ray-tracing code developed in Ref.~\onlinecite{Hojlund24} after the new quadrature rule has been implemented. Shown in red is the raw, unfiltered output from the MGO code, while shown in blue is the result of applying additional high-$k$ filtering to the code output. Specifically, all fluctuations with wavevectors $k > 10 k_\text{max}$ are removed, where $k_\text{max} = 10^{1/3}$ is the maximum wavevector expected in the solution.}
    \label{fig:code_newANGLE}
\end{figure}

The corrected quadrature rule has also been implemented in the MGO code developed in Ref.~\onlinecite{Hojlund24}, which is a complete numerical implementation of MGO. [This means that the exact expressions that comprise \Eq{eq:MGOcusp} are never entered anywhere as input, but are instead computed from first principles.] The result of applying this code to the cusp differential equation \eq{eq:cuspEQ} is shown in \Fig{fig:code_newANGLE}. The raw output of the code (shown in transparent red) is relatively well-behaved; the residual errors are due to spurious poles (Froissart doublets) that get introduced due to the numerical analytical-continuation procedure~\cite{Hojlund24} and might be removable by applying a more advanced version of the analytical-continuation algorithm~\cite{Nakatsukasa18}. 

A simpler remedy is to apply a low-pass filter based on the maximum wavevector expected in the solution. Indeed, since the local wavevector of the solution is already known from the ray trajectory $k(\tau)$, any oscillations with wavevectors much larger than $\text{max}[k(\tau)]$ are spurious and should be filtered out. The results of applying such a low-pass filter to the code output is also shown in \Fig{fig:code_newANGLE}, with visible improvements. It is interesting to note that with the original quadrature rule, the fluctuations in the raw code output were so great that even applying a low-pass filter could not get the solution to be well-behaved near the caustic.

Both the semi-analytical computation and the complete numerical computation of the 1-D cusp caustic achieve acceptable agreement with the exact result. In particular, the semi-analytical computation is able to recover the precise field amplification at the caustic, although the peak is narrower than the exact solution. The reason for this remaining discrepancy is the lack of complex rays in MGO; however, the disagreement is small enough that adding complex rays into the MGO formalism should take lower priority compared to adding new physics capabilities such as mode conversion~\cite{Dodin19}. Also, it is worth mentioning that the MGO solution remains smooth as it enters the caustic because eventually, when the complex ray and real ray become sufficiently close, the numerical algorithm treats them as effectively degenerate. This allows the complex ray contribution to be accounted for in a small neighborhood of the caustic, whereas in exact arithmetic the contribution would only be accounted for exactly at the caustic. 

\section{Modeling `proper shadows' with complex MGO: Fold caustic case study}
\label{sec:airy}

Let us now begin to understand how MGO can be extended to model regions in space where there are no real-valued rays at all, which we call `proper shadows'. As our case study now, we consider the archetypal fold caustic provided by Airy's equation:
\begin{equation}
    \pd{x}^2 E(x) - x E(x) = 0
    ,
    \label{eq:AiryEQ}
\end{equation}

\noindent whose exact (bounded) solution is the Airy function
\begin{equation}
    E_\text{ex}(x) = \airyA(x)
    .
    \label{eq:EXairy}
\end{equation}

\noindent Physically, \Eq{eq:AiryEQ} describes a wave incident from $x < 0$ on a density cutoff at $x = 0$. The propagating wave reflects off the density cutoff and remains at $x < 0$, while the wavefield exponentially decays in the region $x > 0$. 

MGO has already been shown to model correctly the behaviour of \Eq{eq:AiryEQ} for $x < 0$~\cite{Lopez20,Donnelly21}. Now we are interested in modeling $x > 0$. For reference, the standard GO solution for $x > 0$ is given as
\begin{equation}
    E_\text{GO}(x)
    = \frac{
    \exp\left( - \frac{2}{3} x^{3/2} \right)
    }{
        2 \sqrt{\pi} \, x^{1/4}
    }
    .
    \label{eq:GOairy}
\end{equation}

\noindent Clearly, \Eq{eq:GOairy} exponentially decays as $x$ increases, as expected, but there is an erroneous singularity at the caustic $x = 0$ that is not present in the exact solution \eq{eq:EXairy}, nor will it be present in the MGO solution, as we now show.

To generalize \Eq{eq:MGOk!0} for modeling complex-ray contributions with MGO when there are no real-valued rays, we propose to use the following formula based on the general derivation of MGO presented in Ref.~\onlinecite{Lopez22t}:%
\footnote{Specifically, one repeats the derivation of MGO presented in Ref.~\onlinecite{Lopez22t} but now with the underlying $2 \times 2$ symplectic matrix $S$ being complex-valued and unitary; the matrix elements are $S_{22} = S_{11}^* = \dot{x}/|\dot{\Vect{z}}|$ and $S_{21} = - S_{12}^* = - \dot{k}/|\dot{\Vect{z}}|$, with $\Vect{z} = (x, k)$.}
\begin{align}
    \hspace{-2mm}\psi_j(x)
	&=
	\frac{
		c
	}{
		\sqrt{-2\pi i \, [\dot{k}(\tau_j)]^*}
	} 
	\exp\left[
		i \int_0^{\tau_j}
		\dd t \,
		k(t) \dot{x}(t)
	\right]
	\nonumber\\
	&\times
    \int_{\mc{C}_0} \dd \epsilon \,
	\Psi(\epsilon) 
    \exp\left\{ 
        - \frac{i \dot{x}(\tau_j)}{2[\dot{k}(\tau_j)]^*} \epsilon^2 - i K(\tau_j) \epsilon
    \right\}
    ,
    \label{eq:complexMGO}
\end{align}

\noindent where $K(\tau_j)$ is still given by \Eq{eq:rotK}, but with the denominator now involving the sum of complex magnitudes of $\dot{x}$ and $\dot{k}$. Even though this integral mapping is no longer unitary for complex rays (with respect to the standard $L^2$ norm), we anticipate this formula might still be useful because (i) it still obeys the correct asymptotic limit when evaluated far from a caustic~\cite{Lopez22, Lopez22t}, and (ii) it still is a well-defined metaplectic transform when $\Im(\dot{x}/\dot{k}^*) < 0$~\cite{Wolf74,Wolf18}.

The dispersion relation for Airy's equation is
\begin{equation}
    \Symb{D} = k^2 + x = 0
    .
    \label{eq:airyDISP}
\end{equation}

\noindent Consequently, the ray trajectories that solve \Eq{eq:airyDISP} are given by the usual formula
\begin{equation}
	k(\tau) = - \tau, \quad x(\tau) = - \tau^2
    ,
    \label{eq:airyRAYS}
\end{equation}

\noindent but now when $x > 0$, the time parameter is imaginary:
\begin{equation}
	\tau = - i \sqrt{x}
    .
\end{equation}

\noindent The sign of $\tau$ is chosen to have $E_\text{MGO}$ ultimately be a decaying wave; the opposite sign would cause the solution to grow. As in the previous example, there will be only one ray that contributes to the physical solution. It is worth noting that the well-posedness condition for \Eq{eq:complexMGO} constrains how the ray trajectories must be parameterized. For example, although \Eq{eq:airyRAYS} satisfies $\Im(\dot{x}/\dot{k}^*) < 0$, the equally reasonable choice of $k(\tau) = i \tau$, $x(\tau) = \tau^2$ for real $\tau$ does not. Future publications will attempt to elucidate this interplay between complex ray trajectories and complex metaplectic transforms.

Using this ray, all quantities except for $\Psi(\epsilon)$ can be worked out immediately. The result is
\begin{align}
	E_\text{MGO}(x)
	&=
	\frac{
		\exp\left( - \frac{2}{3} x^{3/2} \right)
	}{2\pi}
    \nonumber\\
    &\times
	\int_{\mc{C}_0} \dd \epsilon \,
	\Psi(\epsilon; x) 
	\exp\left[ 
		- \sqrt{x} \, \epsilon^2 
		+ i  \frac{x}{\vartheta(x)} \epsilon
	\right]
 ,
 \label{eq:MGOairyPRE}
\end{align}

\noindent where we have defined the complex magnitude of the ray velocity as $\vartheta(x) \doteq \sqrt{1 + 4 x}$. Next one must compute $\Psi(\epsilon;x)$. The calculation is analogous to that presented in detail in Refs.~\onlinecite{Lopez20,Lopez22t} when deriving the MGO solution of \Eq{eq:AiryEQ} for $x < 0$, so it will not be repeated here for brevity. The final result is
\begin{align}
    \Psi(\epsilon; x)
    &=
    \frac{
        \vartheta(x)
        \, \exp\left[
            \Theta(\epsilon; x)
            + \sqrt{x} \, \epsilon^2 
            - i  \frac{x \epsilon}{\vartheta(x)}
        \right]
    }{
        \left\{
            [\vartheta(x)]^4
            + 8 i \sqrt{x} \, \vartheta(x) \epsilon
        \right\}^{1/4}
    }
    , \\
    \Theta(\epsilon; x)
    &=
    i \frac{[\vartheta(x)]^3}{8 x} \epsilon
	- \frac{ [\vartheta(x)]^2}{4 \sqrt{x}} \epsilon^2
 \nonumber\\
 &
	+ \frac{
		[\vartheta(x)]^6
		- \left\{ 
			[\vartheta(x)]^4 + 8 i \sqrt{x} \, \vartheta(x) \epsilon
		\right\}^{3/2}
	}{96 x^{3/2}}
    .
\end{align}

Having obtained $\Psi(\epsilon; x)$, it is now clear that the integrand of \Eq{eq:MGOairyPRE} decays along the real line as $\epsilon \to \pm \infty$, since it is dominated by the real-valued Gaussian term. It is also important to note that there are no other real-valued saddlepoints besides the one at $\epsilon = 0$; the other saddlepoint is readily derived to be purely imaginary, occurring at $\epsilon = 2 i \sqrt{x}/\vartheta(x)$. One can also verify that the true steepest-descent contour that passes through $\epsilon = 0$ never passes through this additional imaginary saddlepoint, so the integration contour can be deformed to lie simply along the real line. (Away from the caustic, the true steepest-descent curve is well-approximated by the real line anyways.) The final answer for the complex-MGO solution to \Eq{eq:AiryEQ} is therefore 
\begin{align}
	E_\text{MGO}(x)
	&=
	\frac{
		\exp\left( - \frac{2}{3} x^{3/2} \right)
	}{2\pi}
	\nonumber\\
	&\times
	\int_{-\infty}^\infty
	\frac{
		\dd \epsilon \, \vartheta(x) \, 
        \exp\left[
		  \Theta(\epsilon; x)
	   \right]
	}{
		\left\{
			[\vartheta(x)]^4
			+ 8 i \sqrt{x} \, \vartheta(x) \epsilon
		\right\}^{1/4}
	}
 .
 \label{eq:MGOairyCOMPLEX}
\end{align}

\begin{figure}
    \centering
    \includegraphics[width=\linewidth, trim={3mm 13mm 3mm 24mm},clip]{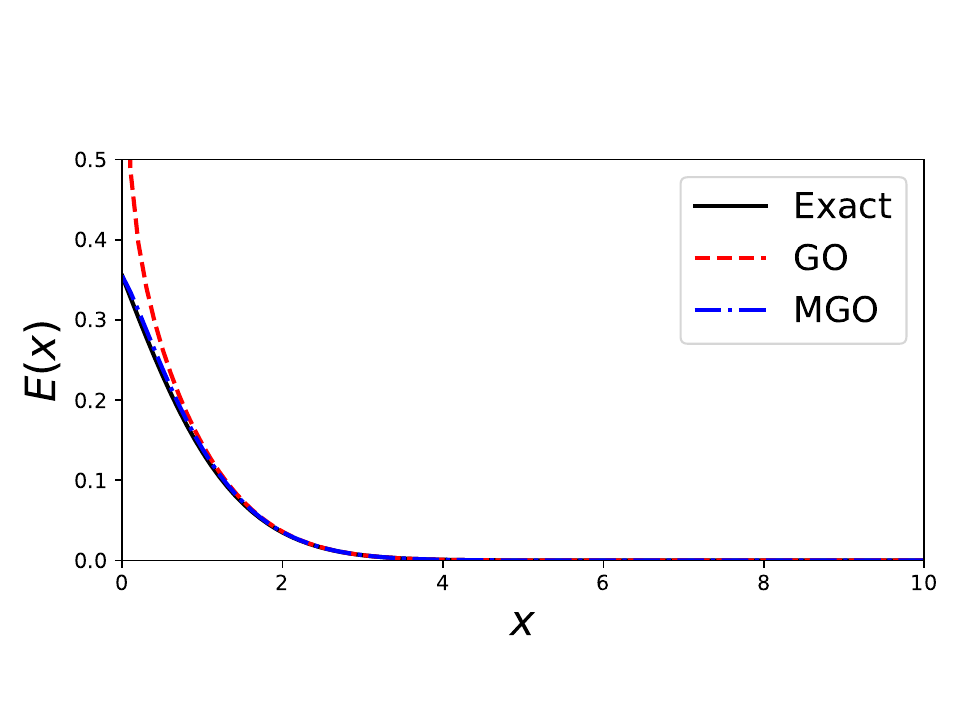}

    \includegraphics[width=\linewidth, trim={3mm 13mm 3mm 24mm},clip]{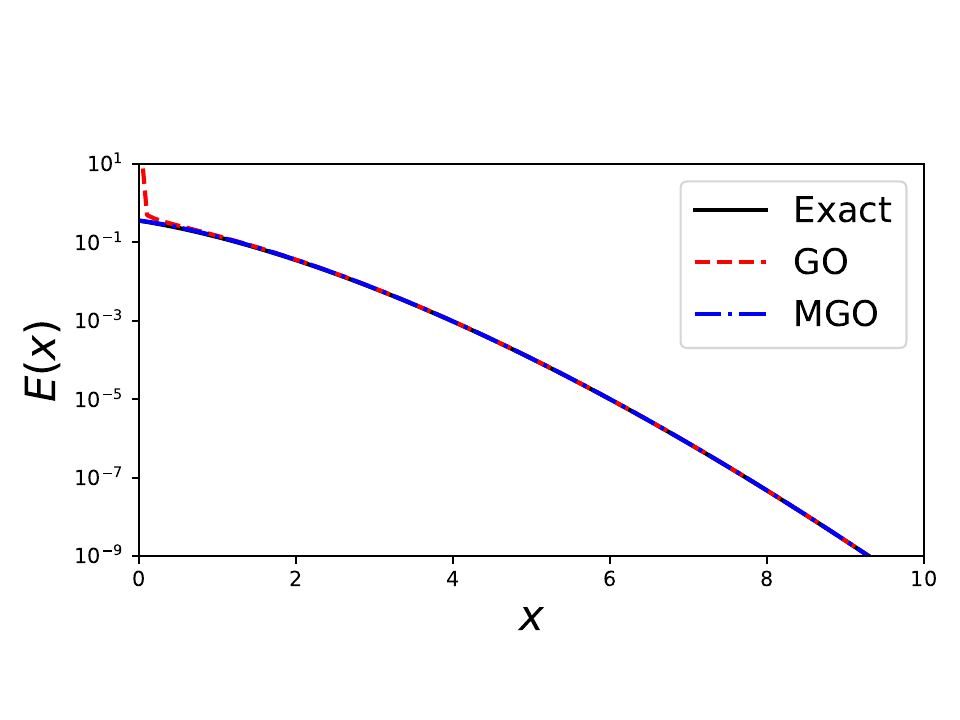}
    \caption{Comparison between the exact solution \eq{eq:EXairy}, the standard GO solution \eq{eq:GOairy}, and the complex-MGO solution \eq{eq:MGOairyCOMPLEX} for the evanescent region of Airy's equation \eq{eq:AiryEQ} shown on a linear scale (top) or a logarithmic scale (bottom).}
    \label{fig:airyCOMPLEX}
\end{figure}

The MGO solution \eq{eq:MGOairyCOMPLEX} is shown in \Fig{fig:airyCOMPLEX} along with the exact solution \eq{eq:EXairy} and the GO solution \eq{eq:GOairy}. Clearly, the MGO solution achieves remarkably high agreement with the exact solution, even when the two curves are compared on a logarithmic scale. This result provides confidence that MGO can be used to model evanescent waves with as equally high accuracy as it can model propagating waves. Lastly, we remark that \Eq{eq:MGOairyCOMPLEX} is actually numerically simpler to compute than the solution for $x < 0$~\cite{Lopez20,Donnelly21} because the integral is performed along the real line. This removes the need to use Gauss--Freud quadrature and allows standard methods to be used instead.

\section{Conclusion}
\label{sec:concl}

MGO is a relatively new framework for modeling caustics within ray-tracing codes that has also recently been implemented in a 1-D code~\cite{Hojlund24}. Thus far, MGO can only model the contribution from real-valued rays that correspond physically to propagating waves. This makes MGO incapable of modeling caustic `shadows' where at least one of the contributing waves is evanescent.

Here we show through two examples how this restriction on MGO might be removed. The first case study is the cusp caustic, whose shadow results from the interference of one real-valued ray and one complex-valued ray. We show that, contrary to what had been previously hypothesized~\cite{Lopez21a}, standard MGO based on real-valued rays can be brought into acceptable agreement with the exact result. This is because, even though the complex ray is formally excluded, its contribution gets naturally included in any numerical implementation when its magnitude becomes comparable to the real-valued ray and the two become nearly coalesced.

The second case study is the fold caustic, whose shadow consists of only a single complex ray corresponding to a purely decaying wavefield. We propose a simple complexification of MGO in which the formalism remains the same but the constituent expressions are allowed to be complex. The resulting complex-MGO solution \eq{eq:MGOairyCOMPLEX} to Airy's equation shows remarkable agreement with the exact result, with no visible difference between the two even when viewed on a logarithmic scale. This formula can be used in MGO codes to include evanescent fields via local asymptotic matching.

This work lays the foundations for a more rigorous derivation of complex-MGO that works when any combination of real-valued and complex-valued rays are present, and lends confidence that such an endeavor will be successful. Such a tool will be useful for modeling higher-order caustics with more complicated shadows, \eg the hyperbolic umbilic caustic that occurs when a focused wave reflects off a plasma density cutoff~\cite{Lopez23}, whose shadow consists both of an illuminated portion and a proper portion. Such a tool will also be useful for describing the tunneling of wavefields through short evanescent regions, as occurs in various mode-conversion schemes to generate the electron Bernstein wave from externally launched electromagnetic waves.

\section*{Acknowledgements}

The work presented here is supported in part by the Carlsberg Foundation, grant CF23-0181. This work has also been carried out within the framework of the EUROfusion Consortium, funded by the European Union via the Euratom Research and Training Programme (Grant Agreement No 101052200 — EUROfusion). Views and opinions expressed are however those of the authors only and do not necessarily reflect those of the European Union or the European Commission. Neither the European Union nor the European Commission can be held responsible for them.



\section*{Disclosures}

The authors declare no conflicts of interest.

\section*{Data availability}

The data that support the findings of this study are available from the corresponding author upon reasonable request.

\appendix

\section{Simplified MGO formulas for semi-analytical computations}
\label{app:simpleMGO}

For rapidly oscillating wavefields that are typically described by ray-tracing methods, one can develop a simplified MGO formalism that makes semi-analytical computations easier. This is accomplished by only considering the phase variation of $\Psi_j(\epsilon)$ within the steepest-descent integral, and then subsequently Taylor-expanding about the saddlepoint until the lowest nonvanishing term arises in the total phase of the integrand. By the strong determinancy of univariate Taylor expansions~\cite{Poston96}, this procedure will yield the correct qualitative behavior at caustics. One ultimately obtains the following simplified expression for the integrand
\begin{align}
	&\hspace{-3mm}
	\Psi_j(\epsilon)
	\exp\left[
		- \frac{
			i \dot{x}(\tau_j)
		}{
			2 \dot{k}(\tau_j)
		} \epsilon^2
		- i K(\tau_j) \epsilon
	\right]
	\nonumber\\
	&\hspace{-3mm}\approx
	\exp\left[
		- \frac{
			i \dot{x}(\tau_j)
		}{
			2 \dot{k}(\tau_j)
		} \epsilon^2
		+ \frac{i K''(\tau_j)}{6} \epsilon^3
		+ \frac{i K'''(\tau_j)}{24} \epsilon^4
		+ ...
	\right]
	,
	\label{eq:MGOapprox}
\end{align}

\noindent with $'$ denoting $\dd/\dd X$, $X$ being the canonically conjugate coordinate to $K$. These derivatives can be readily computed via implicit differentiation of the dispersion relation $\Symb{D}(x,k) = 0$ when written in terms $X$ and $K$~\cite{Lopez20}:
\begin{align}
	\Symb{D}\left( 
		\frac{
			\dot{x} X 
			- \dot{k} K
		}{ \sqrt{ [\dot{x}(\tau_j)]^2 + [\dot{k}(\tau_j)]^2 } }
		,
		\frac{
			\dot{k} X 
			+ \dot{x} K
		}{ \sqrt{ [\dot{x}(\tau_j)]^2 + [\dot{k}(\tau_j)]^2 } }
	, \right) = 0
\end{align}

\noindent Indeed, one obtains
\begin{widetext}
\begin{align}
	\label{eq:Kpp}
	K''(\tau_j)
	&= 
	-
	\frac{
		[\dot{x}(\tau_j)]^2 \pd{x}^2
		\Symb{D}
		+ 2 \dot{x}(\tau_j) \dot{k}(\tau_j) \pd{xk}
		\Symb{D}
		+ [\dot{k}(\tau_j)]^2 \pd{k}^2
		\Symb{D}
	}{ 
		\left\{
			[\dot{x}(\tau_j)]^2
			+ [\dot{k}(\tau_j)]^2 
		\right\}^{3/2}
	}
	,
	\\
	K'''(\tau_j)
	&=
	-\frac{
		[\dot{k}(\tau_j)]^3 \pd{k}^3 \Symb{D}
		+ 3 \dot{x}(\tau_j) [\dot{k}(\tau_j)]^2 \pd{xkk} \Symb{D}
		+ 3 [\dot{x}(\tau_j)]^2 \dot{k}(\tau_j) \pd{xxk} \Symb{D}
		+ [\dot{x}(\tau_j)]^3 \pd{x}^3 \Symb{D}
	}{
		\left\{
			[\dot{x}(\tau_j)]^2
			+ [\dot{k}(\tau_j)]^2 
		\right\}^{2}
	}
	\nonumber\\
	&\hspace{4mm}+ 3
	\frac{
		\dot{x}(\tau_j) \dot{k}(\tau_j) \left( \pd{x}^2 \Symb{D} - \pd{k}^2 \Symb{D} \right)
		+ \left\{ [\dot{k}(\tau_j)]^2 - [\dot{x}(\tau_j)]^2 \right\} \pd{xk} \Symb{D}
	}{
		\left\{
			[\dot{x}(\tau_j)]^2
			+ [\dot{k}(\tau_j)]^2 
		\right\}^{3/2}
	}
	K''(\tau_j)
	,
	\label{eq:Kppp}
\end{align}
\end{widetext}
\noindent and so forth. 

\begin{figure*}
    \centering
    \includegraphics[width=0.19\linewidth,trim={3mm 3mm 3mm 3mm},clip]{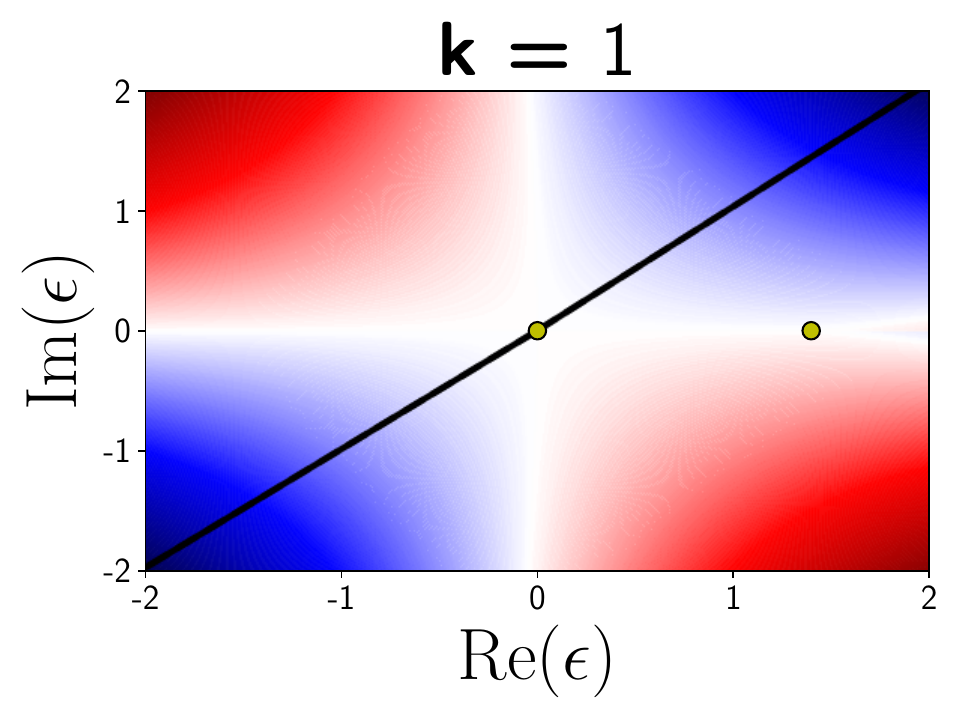}
    \includegraphics[width=0.19\linewidth,trim={3mm 3mm 3mm 3mm},clip]{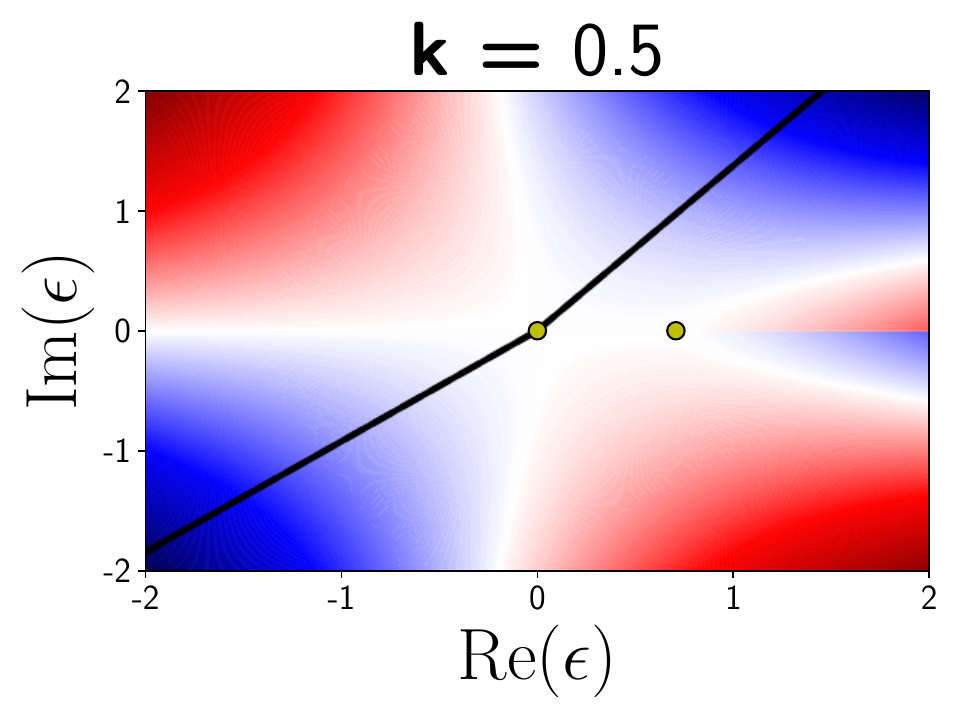}
    \includegraphics[width=0.19\linewidth,trim={3mm 3mm 3mm 3mm},clip]{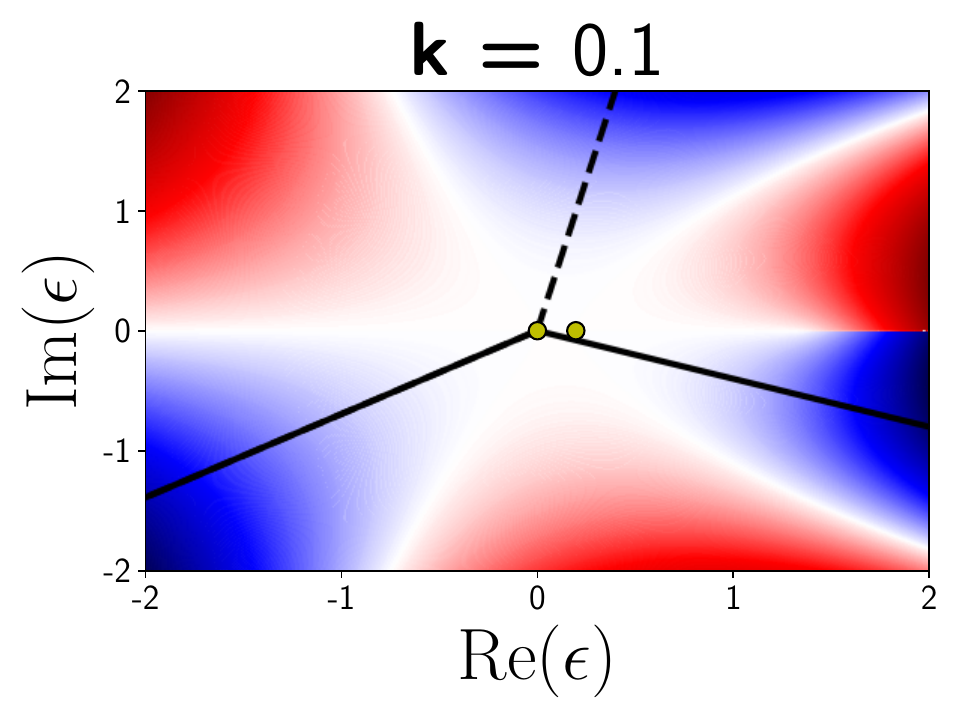}
    \includegraphics[width=0.19\linewidth,trim={3mm 3mm 3mm 3mm},clip]{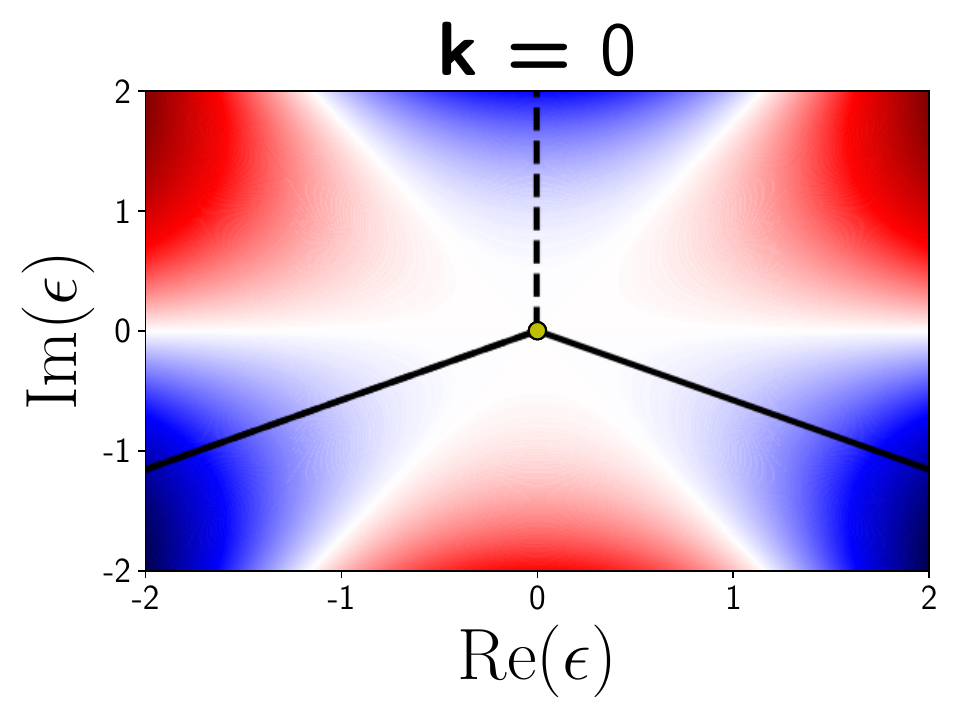}
    \includegraphics[width=0.19\linewidth,trim={3mm 3mm 3mm 3mm},clip]{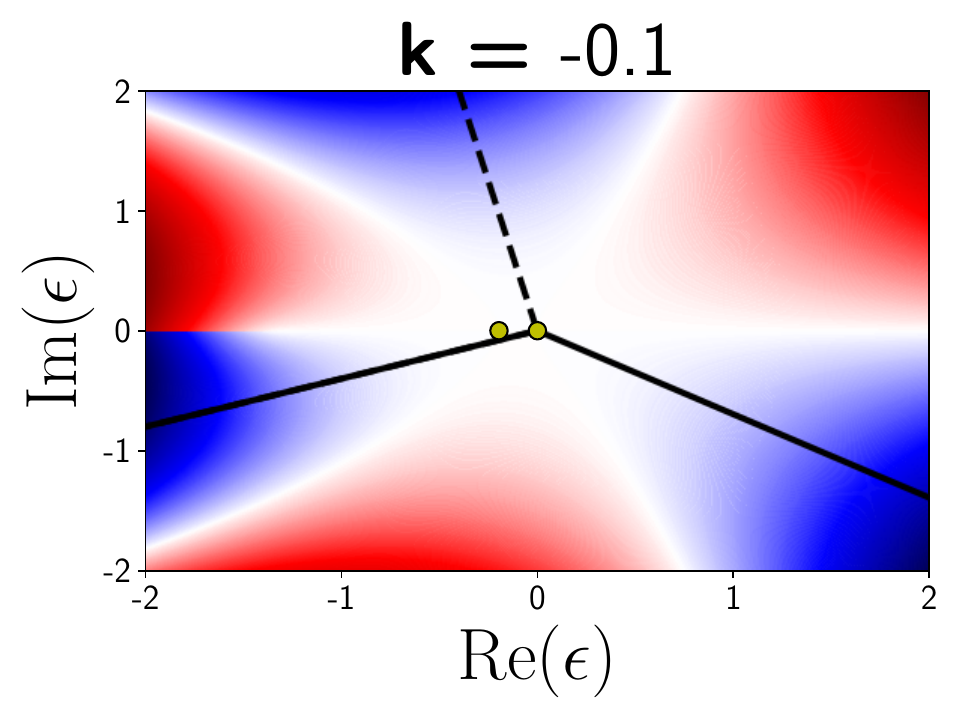}

    \vspace{2mm}
    \includegraphics[width=0.19\linewidth,trim={3mm 3mm 3mm 3mm},clip]{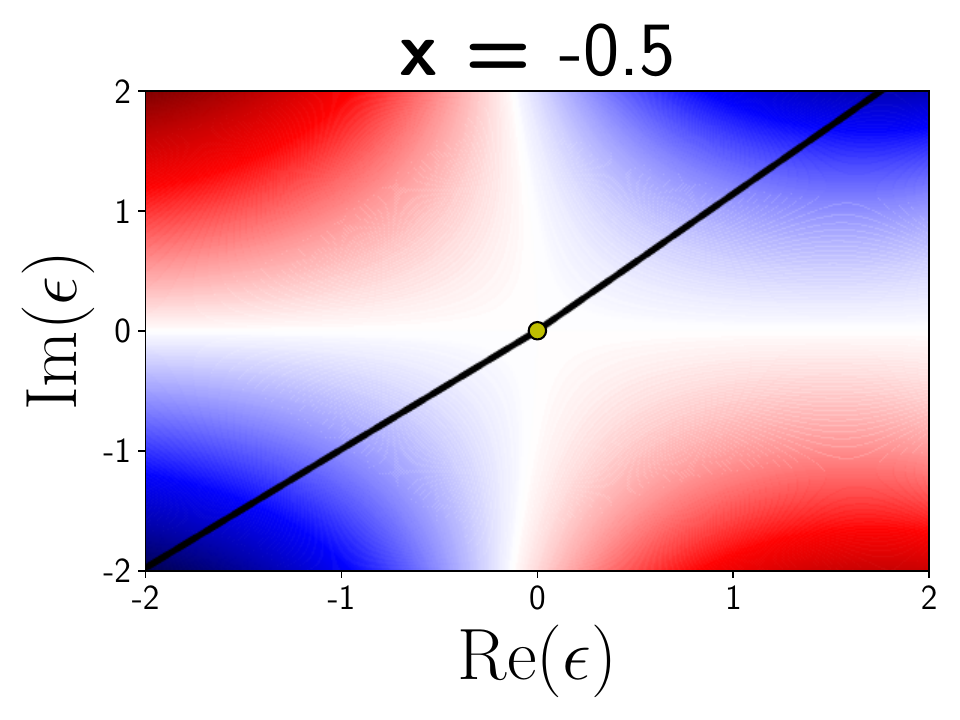}
    \includegraphics[width=0.19\linewidth,trim={3mm 3mm 3mm 3mm},clip]{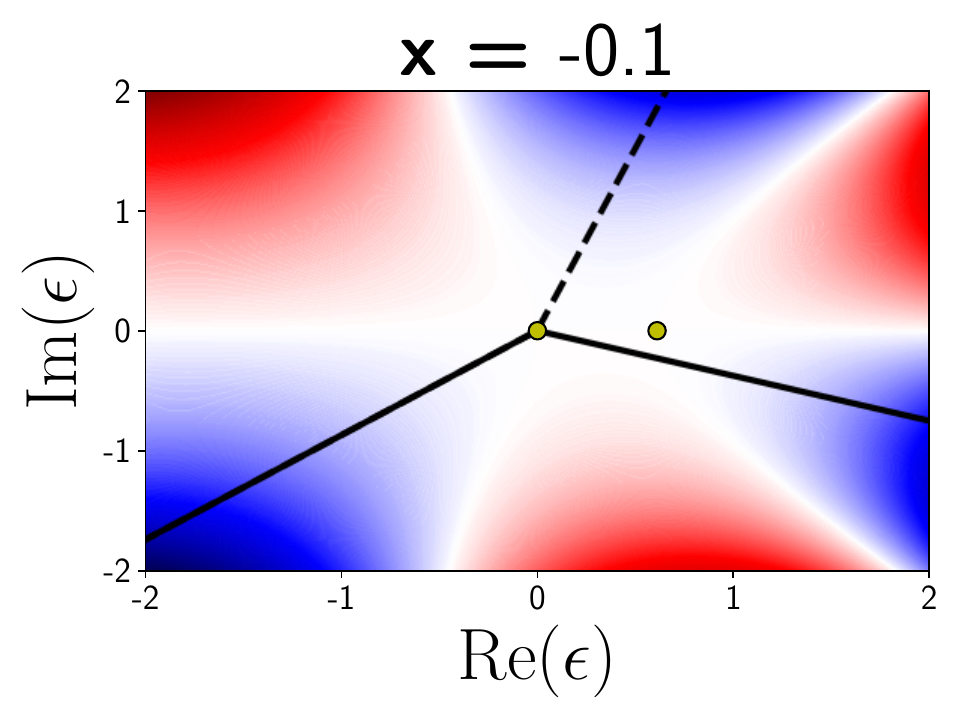}
    \includegraphics[width=0.19\linewidth,trim={3mm 3mm 3mm 3mm},clip]{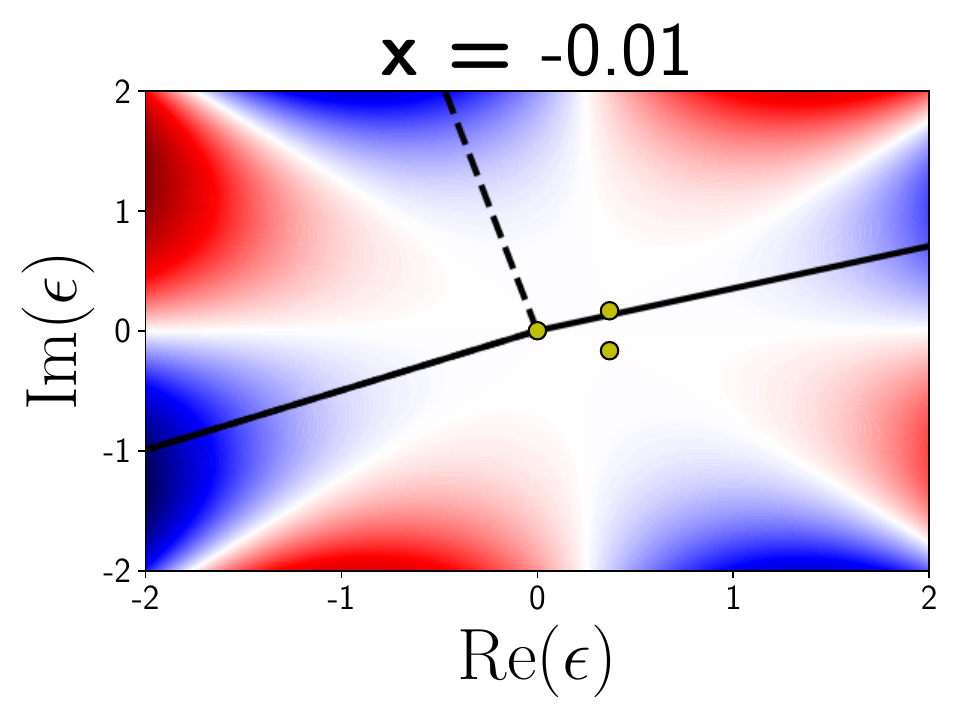}
    \includegraphics[width=0.19\linewidth,trim={3mm 3mm 3mm 3mm},clip]{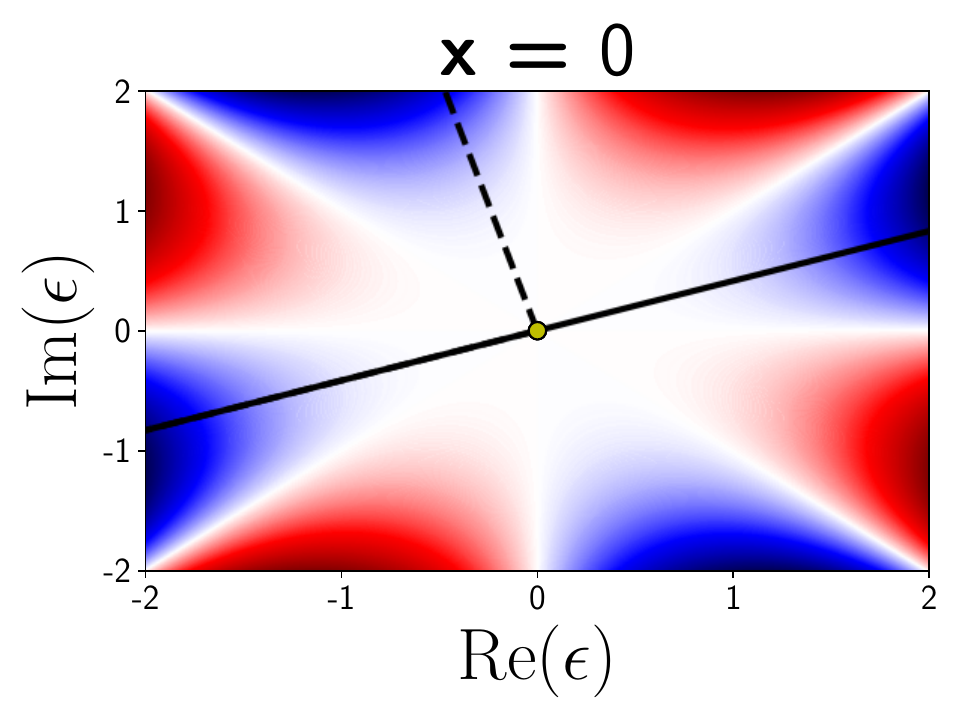}
    \includegraphics[width=0.19\linewidth,trim={3mm 3mm 3mm 3mm},clip]{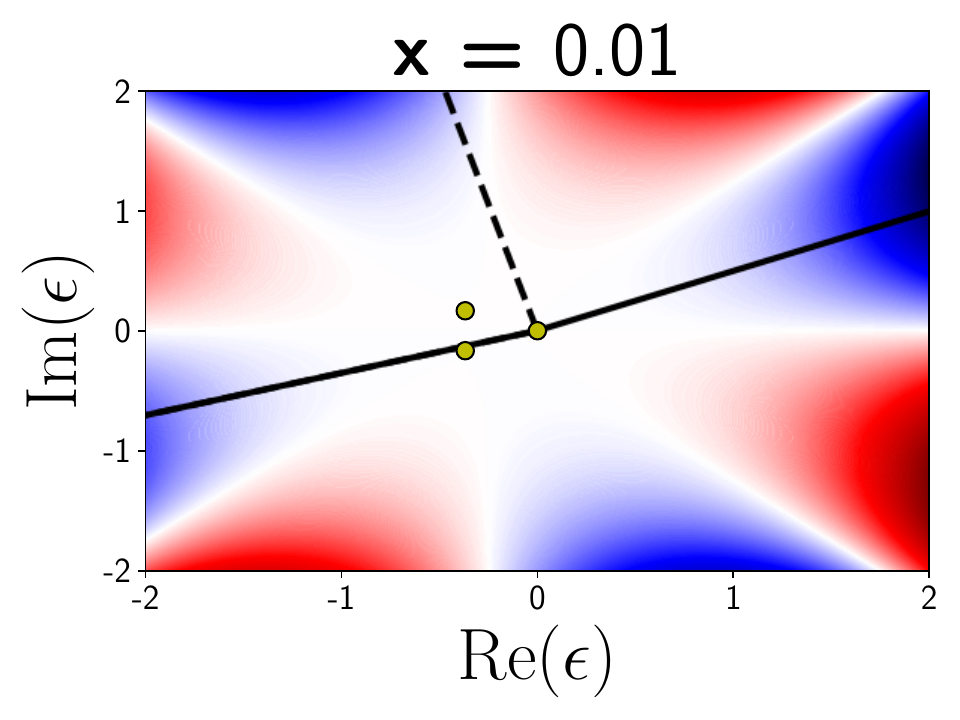}
    \caption{\textbf{(Top row)} Evolution of the identified steepest-descent contours for a fold caustic (\Sec{sec:airy}) using either the old angle bias (black dashed line), which preferentially selects contours that are minimally changed from the previous timestep, or the new angle bias (black solid line), which preferentially selects contours that are minimally changed from the real line. The caustic occurs at $k = 0$, where $k$ is the local wavevector that evolves in space as $k = \pm \sqrt{-x}$. The sign of $k$ determines whether it corresponds to the incident or reflected wave. Note that the phase function used here corresponds to the illuminated region of the caustic, and is derived in Ref.~\onlinecite{Lopez20}. \textbf{(Bottom row)} Same as top, except for the cusp caustic (\Sec{sec:cusp}). Here the caustic occurs at $x = 0$, with $x$ being the spatial location of the wavefield. Note, however, that these series of figures use the simplified MGO phase function described in \App{app:simpleMGO} rather than the exact expressions used in \Sec{sec:cusp}. In all plots, the color indicates the magnitude of the integrand within \Eq{eq:MGOk!0}, with blue being smaller values and red being larger ones; hence a valid steepest-descent contour connects one of the saddlepoints (denoted by a yellow dot) with one of the blue regions.}
    \label{fig:contEVO}
\end{figure*}

However, one should be careful when using \Eq{eq:MGOapprox} since the root structure of the integrand phase is not necessarily preserved; consequently, the steepest-descent topology may not be preserved either. This is seen in \Fig{fig:contEVO}, which shows how the steepest-descent topology evolves for the cusp caustic (\Sec{sec:cusp}) using \Eq{eq:MGOapprox} with
\begin{align}
	K''(\tau_j)
	=
	\frac{
		6 \tau_j
	}{ 
		\left(
			9 \tau_j^4 + 1
		\right)^{3/2}
	}
	, \quad
	K'''
	=
	\frac{
		6 - 270 \tau_j^4
	}{
		\left(
			9 \tau_j^4 + 1
		\right)^{3}
	}
	.
\end{align}

\noindent It is clear that for $|x| > 0.02$ there are three real saddlepoints. This is actually an artifact of the Taylor expansion, and the true phase function has only one real saddlepoint for all $x$. That said, the corrected quadrature-angle bias (\App{app:correctANGLE}) minimizes the impact of these spurious saddlepoints; using \Eq{eq:MGOapprox} with the corrected angle bias gives figures that are nearly identical with those presented in \Fig{fig:cusp_new} based on the exact expressions.

\section{Corrected quadrature-angle bias for MGO}
\label{app:correctANGLE}

\subsection{Context: Numerical steepest-descent integration}

Both the exact MGO formula \eq{eq:MGOk!0} and the approximate form given by \Eq{eq:MGOapprox} involve a highly oscillatory integral of the form $\int \dd \epsilon A(\epsilon) e^{i \theta(\epsilon)}$, where $A$ is a slowly varying amplitude and $\theta$ is a rapidly varying phase. In Ref.~\onlinecite{Donnelly21} it was proposed to compute this integral along straight-line approximations to the steepest-descent curves via Gauss--Freud quadrature, which is exact for quadratic phase functions; as such, it is exact when evaluated away from a caustic and converges to the correct answer when evaluated at a caustic as the quadrature order is increased. The algorithm contains two key steps: first, the correct rotation angle to obtain the desired integration contour from the real line must be identified, then second, the appropriate quadrature rule must be applied. We shall discuss the first step in the following section; here we briefly outline the second step for completeness. Note that the following discussion assumes the appropriate functions can be obtained in the complex plane; see Ref.~\onlinecite{Hojlund24} for how to apply these results for data that is only sampled along the real line.

Suppose we have already obtained the angles $\phi_-$ and $\phi_+$ that the incoming and outgoing portions of the integration contour passing through $\epsilon = 0$ make with respect to the real line. We transform from $\epsilon$ to the natural parameterization $\ell$ of the integration contour as
\begin{equation}
    \epsilon(\ell; \tau_j) \approx
    |\ell| \times
    \left\{
        \begin{array}{cc}
            \exp\left[i \phi_-(\tau_j) \right], & \ell \le 0 \\
            \exp\left[i \phi_+(\tau_j) \right], & \ell > 0
        \end{array}
    \right.
    .
\end{equation}

\noindent We have included a functional dependence on $\tau_j$ since in general the angles $\phi_\pm$ will change along a ray.

Having rotated the integration contour to lie in the directions of steepest descent, one now applies the Gauss--Freud quadrature rule~\cite{Donnelly21}:
\begin{subequations}
    \label{eq:GFquad}
    \begin{align}
        \int \dd \epsilon \, I(\epsilon; \tau_j) &=
        \sum_{m = 1}^n w_m
        \left[
            h_+(\ell_m; \tau_j) - h_-(\ell_m; \tau_j)
            \nullFrac
        \right]
        e^{\ell_m^2}
        , \\
        h_\pm(\ell; \tau_j) &\doteq
        I\left[
            \frac{\ell e^{i \phi_\pm(\tau_j)}}{\sqrt{s_\pm(\tau_j)}}
            ; \tau_j
        \right]
        \frac{e^{i \phi_\pm(\tau_j)}}{\sqrt{s_\pm(\tau_j)}}
        ,
    \end{align}
\end{subequations}

\noindent where $I(\epsilon; \tau_j) = A(\epsilon; \tau_j) e^{i \theta(\epsilon; \tau_j)}$ denotes the integrand. Here, $\{ w_j \}$ and $\{ \ell_j \}$ are the quadrature weights and nodes for the Freud polynomials, respectively. Tables of their values are provided in \Refs{Donnelly21,Steen69}. Also, $s_\pm(\tau_j)$ are the re-scaling factors for each branch of the integration contour, defined as the squared e-folding length of the integrand along the rotated contour. The introduction of $s_\pm$ is needed to allow the quadrature rule to be written in terms of a single fiducial weight function.

The quadrature rule \eq{eq:GFquad} is then applied to evaluate either \Eq{eq:MGOk!0} or \Eq{eq:MGOapprox} at a given point $\tau_j$ along a ray. The quadrature angles $\phi_\pm$ evolve as one moves along a ray, however; one can compute them at each step via the root-finding procedure
\begin{equation}
    \phi_\pm(\tau_j) = \left.\arg\left( \epsilon_\pm \right) \, \nullFrac\right| \, i \theta(\epsilon_\pm; \tau_j) = -1
    .
    \label{eq:PHIfit}
\end{equation}

\noindent Unfortunately, the solutions to the root-finding will not be unique if the saddlepoint is degenerate (in actuality, or simply numerically), as occurs at caustics. Hence, some additional constraint is required to break the degeneracy and allow a numerical code to operate without supervision. We shall now discuss this issue in more detail.

\subsection{Old versus new angle bias}

The solution proposed by Ref.~\onlinecite{Donnelly21} is to choose the angles such that the difference between the chosen angle and the angle used in the previous step is minimal. Thus, one biases the root-finding search around the previous angle. The logic behind this proposal is that the steepest-descent contours evolve smoothly away from a caustic. However, they can undergo discontinuous changes at/across caustics. Hence, if one is not careful, ray-tracing codes based on this proposal can fail completely after encountering the first caustic in a simulation%
\footnote{Advanced techniques can be used to overcome this difficulty, at least for the simplest class of caustics - see Ref.~\onlinecite{Hojlund24}}. %
This is illustrated in \Fig{fig:contEVO}, which shows how biasing the angle towards the previously selected angle eventually causes the wrong angles to be selected as one encounters a fold and a cusp caustic. (The integration contours should have a definite parity about the caustic for both cases shown.) This implies that the Gauss--Freud quadrature rule will not converge to \Eq{eq:MGOk!0} at a caustic as the quadrature order is increased, simply because the integration contour used is not correct.

Also shown in \Fig{fig:contEVO} are the angles chosen by a new bias that we propose here: rather than minimizing the difference with the previous angle, one should instead minimize the difference with the real line (which is more like how steepest-descent integration is often done). In other words, the outgoing angle $\phi_+$ is biased towards
\begin{subequations}
    \begin{equation}
        \phi_+ = 0
        ,
    \end{equation}

    \noindent and the incoming angle $\phi_-$ is biased towards
    \begin{equation}
        \phi_- = \pm \pi
        .
    \end{equation}
\end{subequations}

\noindent (The sign convention depends on code details.) As illustrated in the figure, this angle bias selects the correct contours at caustics, and also does not fail after one crosses a caustic. 

Implementing the new angle bias is particularly simple for existing codes based on the old angle bias, perhaps by changing only a single command. The numerical performance of such codes is then greatly improved, not only by enabling the codes to progress beyond caustics smoothly and without user intervention, but also by decoupling the contour selection from the stepsize used in the ray-tracer. Indeed, by only looking at angles near the previous one, the old angle bias inadvertently placed strong constraints on the stepsize so that the contour does not change appreciably from one time step to the next. With the new angle bias, each contour selection is independent from previous ones, so one can now take arbitrarily large timesteps if desired.

\bibliography{Biblio.bib}
\bibliographystyle{apsrev4-1}
\end{document}